\documentclass[%
 reprint,
 amsmath,amssymb,
 aps,
]{revtex4-1}

\usepackage[pdftex]{graphicx,color}% Include figure files
\usepackage{dcolumn}% Align table columns on decimal point
\usepackage{bm}% bold math

\usepackage{comment}% for edit
\usepackage{ulem}% for edit

\begin{document}

\preprint{APS/123-QED}

\title{Effect of spin fluctuations on superconductivity in V and Nb: a first-principles study}% Force line breaks with \\
%\thanks{A footnote to the article title}%

\author{Kentaro Tsutsumi$^{1}$}
\author{Yuma Hizume$^{1}$}
\thanks{yuma.hizume@phys.s.u-tokyo.ac.jp}
\author{Mitsuaki Kawamura$^{2}$}%
\author{Ryosuke Akashi$^{1}$}
\author{Shinji Tsuneyuki$^{1,2}$}
\affiliation{$^1$Department of Physics, the University of Tokyo, Hongo, Bunkyo-ku 113-8656, Japan}
\affiliation{$^2$Institute for Solid State Physics, the University of Tokyo, Kashiwa, 277-8581, Japan}
% \email{Second.Author@institution.edu}
%\affiliation{%
% Authors' institution and/or address\\
% This line break forced with \textbackslash\textbackslash
%}%

\date{\today}% It is always \today, today,
             %  but any date may be explicitly specified

\begin{abstract}
We study the superconductivity in typical $d$-band elemental superconductors V and Nb with the recently developed non-empirical computational scheme based on the density functional theory for superconductors. The effect of ferromagnetic fluctuation (paramagnon) on the superconducting transition temperature ($T_{\rm c}$), which in principle suppress the $s$-wave superconducting pairing, is quantified without any empirical parameter. We show that the strong paramagnon effect cancels the $T_{\rm c}$-enhancing effects of the phonon-mediated pairing and dynamical screened Coulomb interaction.  
%\begin{description}
%\item[Usage]
%Secondary publications and information retrieval purposes.
%\item[PACS numbers]
%May be entered using the \verb+\pacs{#1}+ command.
%\item[Structure]
%You may use the \texttt{description} environment to structure your abstract;
%use the optional argument of the \verb+\item+ command to give the category of each item. 
%\end{description}
\end{abstract}

\pacs{Valid PACS appear here}% PACS, the Physics and Astronomy
                             % Classification Scheme.
%\keywords{Suggested keywords}%Use showkeys class option if keyword
                              %display desired
\maketitle

%\tableofcontents

\section{\label{sec:intro}INTRODUCTION}
Since its discovery~\cite{Onnes1911}, the microscopic mechanism in superconductors has been one of the central topics of theoretical physics. At first, Bardeen, Cooper, and Schrieffer (BCS)~\cite{BCS1957} 
successfully constructed the microscopic theory of phonon-driven superconductors.
Later, the dynamical structure of the electron-phonon interaction has been studied employing the 
Green's function formalism~\cite{Migdal1958,Nambu1960,Eliashberg1960,Morel1962,Schrieffer1964,Scalapino1966} 
and this theory is called the Migdal-Eliashberg (ME) theory.
However, there are some problems in the practice of the ME theory. First, there is an empirical parameter $\mu^{\ast}$
which represents the effect of the electronic Coulomb interaction suppressing the superconducting pairing. 
As long as there is an empirical parameter, 
the ME theory is not appropriate for a quantitative prediction of $T_{\rm c}$.
Another reason is that, when we introduce $\mu^{\ast}$, it is implicitly assumed that electronic
Coulomb repulsion suppresses the pairing of electrons. Within this restriction, we cannot treat the
pairing originating from the repulsive interaction, e.g., plasmon and spin-fluctuations mechanisms.

In order to avoid the empirical treatment, an extension of the density functional theory for superconductors (SCDFT)
has been developed~\cite{Oliveira1988}. Based on this theory, a numerical scheme to calculate
$T_{\rm c}$ without introducing any empirical parameter has been implemented recently~\cite{Luders2005}.
It has been shown that this scheme reproduces the experimental $T_{\rm c}$ of various kinds of 
phonon-mediated conventional superconductors, such as elemental metals~\cite{Marques2005}, $\rm MgB_2$~\cite{Floris2005} and $\rm CaC_6$~\cite{Sanna2007} within an accuracy of a few Kelvin even if only 
the electron-phonon interaction and static effect of Coulomb interaction are taken into account.
%There also have been some applications of this scheme to other materials\cite{Profeta2006Pressure,Bersier2009CaBeSi,RA2012,RA2013alkali}

This numerical scheme has been extended to include the plasmon effect~\cite{RA2013}, namely, 
the dynamical effect of the Coulomb interaction. Originally, the plasmon mechanism in 2D and 
3D homogeneous electron gas has been studied with the Green's function formalism~\cite{Takada1978}, 
and $T_{\rm c}$ of some materials has been calculated~\cite{Takada1980,Takada1982,Takada2009}. It has been confirmed that the effect of plasmon cooperates with phonons in superconducting pairing with the DFT-based scheme. 

Furthermore, an extension to include the effect of spin fluctuations 
has also been developed in 2014~\cite{Essenberger2014}
and applied to explore the superconducting pairing driven by the spin fluctuations
in the Fe-based superconductors~\cite{Essenberger2016}.
On the other hand, the effect of spin fluctuations is also crucial in conventional phonon-driven superconductors.
If the parallel spin alignment is favored due to the exchange effect, the singlet pairing should be suppressed~\cite{Parks1969}.
Such effect of spin fluctuations in conventional superconductors has been studied by employing the particle-hole $t$-matrix within the ME theory, including parameters~\cite{Berk1966,Schrieffer1968}.
However ,the case studies on the reduction of $T_{\rm c}$ with fully nonempirical calculations are yet few\footnote{In fact, there has been an extension to include the effect of spin fluctuations~\cite{Wierzbowska2005}.
However, this extension includes a parameter representing the electron-paramagnon coupling constant, and therefore, we do not employ this one.}.

Recently, some of the authors conducted exhaustive calculations of $T_{\rm c}$ with the effect of spin fluctuations for elemental metals~\cite{Kawamura2020}. In the present study, we delve more into how it affects $T_{\rm c}$ with a detailed analysis of selected systems; elemental V, Nb, and Al. Within the ME theory, Rietschel and Winter have stated that the $T_{\rm c}$'s in V and Nb could be overestimated without the spin effect~\cite{Rietschel1979}, whereas Marques {\it et al}., has demonstrated an accurate estimate for Nb and Al with SCDFT without spin fluctuations~\cite{Marques2005}. We carefully study how $T_{\rm c}$ of each material is affected by the plasmon and spin fluctuations 
and infer a general principle in what kind of material spin fluctuations reduce $T_{\rm c}$ strongly.

\section{METHOD}
The central equation of SCDFT to calculate $T_{\rm c}$ from first principles is the following 
gap equation 
\begin{equation}
	\Delta_{n\bm{k}} = -\mathcal Z_{n\bm{k}}\Delta_{n\bm{k}}
	- \frac{1}{2}\sum_{n'\bm{k'}}\mathcal K_{n\bm{k}n'\bm{k'}}
	\frac{\tanh[(\beta/2)E_{n'\bm{k'}}]}{E_{n'\bm{k'}}}\Delta_{n'\bm{k'}},
	\label{eq:SCDFTgap}
\end{equation}
where $\beta$ is the inverse temperature, $\Delta$ is the gap function, $n$ and $\bm k$ indicate
the band index and the crystal momentum, $E_{n\bm k}$ is defined as 
$E_{n\bm k}=\sqrt{\xi_{n\bm k}^2 + |\Delta_{n\bm k}|^2}$ and $\xi_{n\bm k}=\varepsilon_{n\bm k}-\mu$
is the Kohn-Sham eigenenergy measured from the chemical potential $\mu$.
In the kernels $\mathcal Z$ and $\mathcal K$, we include the effect of electron-phonon interaction
($\mathcal{Z}^{\rm ph}$, $\mathcal{K}^{\rm ph}$),
electron-electron Coulomb interaction ($\mathcal{K}^{\rm el}$),
and the effective interaction driven by spin fluctuations ($\mathcal{Z}^{\rm SF}$, $\mathcal{K}^{\rm SF}$) as
\begin{align*}
	\mathcal{Z} &= \mathcal{Z}^{\rm ph} + \mathcal{Z}^{\rm SF}\\
	\mathcal{K} &= \mathcal{K}^{\rm ph} +\mathcal{K}^{\rm el}+ \mathcal{K}^{\rm SF}.
\end{align*}
We briefly review the expression of each contribution below.

The kernels $\mathcal{Z}^{\rm ph}$ and $\mathcal{K}^{\rm ph}$ originating from the electron-phonon interaction are approximated as \cite{Marques2005}
\begin{equation}
	\begin{split}
	\label{eq:Kphdef}
	&\mathcal K^{\rm ph}_{n\bm k n' \bm k'} =
	\frac{2}{\tanh[(\beta/2)\xi_{n\bm k}]\tanh[(\beta/2)\xi_{n'\bm k'}]}
	\frac{1}{N(0)} \\
	& \times \int d\omega \alpha^2F(\omega)
	[I(\xi_{n\bm k},\xi_{n'\bm k'},\omega)-I(\xi_{n\bm k},-\xi_{n' \bm k'},\omega)],
\end{split}
\end{equation}
\begin{equation}
	\begin{split}
	\label{eq:Zphdef}
	\mathcal Z^{\rm ph}_{n\bm k} &= -\frac{1}{\tanh[(\beta/2)\xi_{n\bm k}]}
	\int^{\infty}_{-\mu}d\xi' \int d\omega 	\alpha^2F(\omega)  \\
	&\quad \times
	[J(\xi_{n\bm k},\xi', \omega)+ J(\xi_{n\bm k}, \xi',\omega)]. 
\end{split}
\end{equation}
Here $N(\xi)$ is the density of states,
and $\alpha^2F(\omega)$ is called the Eliashberg function and defined as
\begin{equation}
	\alpha^2F(\omega) = \frac{1}{N(0)}\sum_{\substack{nn'\bm k \\ \bm q \nu}}
	\delta(\xi_{n\bm {k+q}})\delta(\xi_{n'\bm k})\delta(\omega-\omega_{\nu \bm{q}})
	|g^{n\bm {k+q},n'\bm k}_{\nu \bm{q}}
	|^2,
	\label{eq:Eliashbergfunc}
\end{equation}
where $g$ is the electron-phonon coupling constant defined as
\begin{equation}
	g^{n\bm {k+q},n'\bm k}_{\nu \bm{q}} = \int d^3r 
	\varphi^{\ast}_{n\bm{k+q}} \delta V^{\rm KS}_{\lambda\bm q}(\bm r)
	\varphi_{n'\bm k}(\bm r),
	\label{eq:elphcoupling}
\end{equation}
$\varphi$ is the Kohn-Sham wavefunction, and $\delta V^{\rm KS}_{\lambda\bm q}$ is the gradient of the electronic Kohn-Sham potential with respect to the collective coordinates of nuclei.
The functions $I$ and $J$ are defined in Ref.~\onlinecite{Luders2005}. 

The contribution of electronic Coulomb interaction to the kernel $\mathcal K^{\rm el}$ is written 
as~\cite{RA2013}
\begin{equation}
	\begin{split}
	\label{eq:Keldyn}
  \mathcal K^{\rm el}_{n\bm kn'\bm k'} &= \lim_{ \{\Delta_{n\bm k}\}\to 0}
	\frac{1}{\tanh[(\beta/2)E_{n\bm k}]}\frac{1}{\tanh[(\beta/2)E_{n'\bm k'}]}
	\frac{1}{\beta^2} \\
	&\quad \times
	\sum_{\omega_1\omega_2}F_{n\bm k}( {\rm i}\omega_1)F_{n'\bm k'}( {\rm i}\omega_2)
	W_{n\bm k n' \bm k'}[{\rm i}(\omega_1-\omega_2)],
\end{split}
\end{equation}
where $F_{n\bm k}({\rm i}\omega) = 
\frac{1}{ {\rm i} \omega + E_{n\bm k}}-\frac{1}{ {\rm i} \omega - E_{n\bm k}}$
and $\omega$ means the fermionic Matsubara frequency. $W_{n\bm kn'\bm k'}({\rm i}\nu)$ is defined as
\begin{equation}
	\begin{split}
	\label{eq:dynWnknk}
	W_{n\bm kn'\bm k'}({\rm i}\nu) &= 
	\iint d^3r d^3r' \varphi^*_{n\bm k}(\bm r)\varphi^*_{n'\bm k'}({\bm r'})
	v^{\rm scr}(\bm r, {\bm r'},{\rm i}\nu) \\
	&\quad \times \varphi_{n\bm k}({\bm r'})\varphi_{n'\bm k'}(\bm r),
\end{split}
\end{equation}
where $v^{\rm scr}$ is the screened Coulomb interaction, and $\nu$ is the bosonic Matsubara frequency.
We apply the adiabatic local density approximation (ALDA)~\cite{Zangwill1980} to 
$v^{\rm scr}$ and then it is written as
\begin{equation}
	\begin{split}
	\label{eq:vscr}
	&v^{\rm scr}(\bm r,\bm r', {\rm i}\nu) = \frac{1}{|\bm r - \bm r'|}  \\
	&\quad +
	\iint d^3r_1 d^3r_2 \left( \frac{1}{|\bm r - \bm r_1|}+
	f_{\rm xc}(\bm r,\bm r_1) \right) 
	\chi(\bm r_1, \bm r_2, {\rm i}\nu)\\
	&\quad \times \frac{1}{|\bm r_2 - \bm r'|},
\end{split}
\end{equation}
where $\chi$ is the polarization function obtained by the following equation
\begin{equation}
	\begin{split}
	\label{eq:chieq}
	&\chi(\bm r, \bm r', {\rm i}\nu) = \chi_0(\bm r, \bm r', {\rm i}\nu) \\
	&\quad + 
	\iint d^3r_1d^3r_2\chi_0(\bm r, \bm r_1, {\rm i}\nu) 
	\left(
	\frac{1}{|\bm r_1 - \bm r_2|} + f_{\rm xc}(\bm r_1, \bm r_2)
	\right) \\
	&\quad \quad \times
	\chi(\bm r_2, \bm r', {\rm i}\nu).
\end{split}
\end{equation}
$\chi_0$ represents the independent-particle polarization function in the Kohn-Sham system
\begin{equation}
	\begin{split}
	\label{eq:chi0}
	\chi_0(\bm r, \bm r', {\rm i}\nu) &= \sum_{n\bm k n' \bm k'}
	\frac{\theta(-\xi_{n\bm k})-\theta(-\xi_{n'\bm k'})}{\xi_{n\bm k}-\xi_{n'\bm k'}+{\rm i}\nu} \\
	&\quad \times
	\varphi^\ast_{n\bm k}(\bm r)\varphi^\ast_{n'\bm k'}(\bm r')\varphi_{n\bm k}(\bm r')
	\varphi_{n'\bm k'}(\bm r),
\end{split}
\end{equation}
where $\theta(\xi)$ is the step function.
$f_{\rm xc}$ is defined as the functional derivative of the exchange-correlation energy
$E_{\rm xc}$ with respect to the electronic density $\rho$:
\begin{equation}
	f_{\rm xc}(\bm r, \bm r') = 
	\frac{\delta^2E_{\rm xc}}{\delta\rho(\bm r)\delta\rho(\bm r')}.
	\label{eq:Excrhorho}
\end{equation}
In practice, we decompose $\mathcal K^{\rm el}$ into static ($\mathcal K^{\rm el,stat}$)
and dynamical ($\mathcal{K}^{\rm el,dyn}$) terms
\begin{equation}
	\begin{split}
	\label{eq:Keldecomp}
  \mathcal K^{\rm el}_{n\bm k n'\bm k'} &= \mathcal K^{\rm el,stat}_{n\bm k n' \bm k'} + 
  \mathcal K^{\rm el,dyn}_{n\bm k n'\bm k'}, \\
  \mathcal K^{\rm el,stat}_{n\bm k n'\bm k'} &= W_{n\bm k n' \bm k'}({\rm i}\nu=0), \\
  \mathcal K^{\rm el,dyn}_{n\bm k n'\bm k'} &= \lim_{ \{\Delta_{n\bm k}\}\to 0}
	\frac{1}{\tanh[(\beta/2)E_{n\bm k}]}\frac{1}{\tanh[(\beta/2)E_{n'\bm k'}]}
	\frac{1}{\beta^2}\\
	&\quad \times
	\sum_{\omega_1\omega_2}F_{n\bm k}( {\rm i}\omega_1)F_{n'\bm k'}( {\rm i}\omega_2)\\
	&\quad \quad \times
	\{W_{n\bm k n' \bm k'}[{\rm i}(\omega_1-\omega_2)] - W_{n\bm k n' \bm k'}({\rm i}\nu=0)\},
\end{split}
\end{equation}
where $\mathcal K^{\rm el,dyn}$ represents the contribution of plasmon effect to the electron-electron kernel. The summation with respect to $\omega_1$ in Eq. (\ref{eq:Keldecomp}) is calculated analytically and we obtain
\begin{widetext}
\begin{equation}
\begin{split}
	\label{eq:Ksfdyn}
    \mathcal K^{\rm el,dyn}_{n\bm k n'\bm k'} &= 
	\frac{1}{\tanh[(\beta/2)\xi_{n\bm k}]}\frac{1}{\tanh[(\beta/2)\xi_{n'\bm k'}]}
	\left[
	{\rm sgn}(\xi_{n\bm k}-\xi_{n'\bm k'})
	\left[
	f(\xi_{n\bm k})-f(\xi_{n'\bm k'})
	\right]
	L_{n\bm k n'\bm k'}(|\xi_{n\bm k}-\xi_{n'\bm k'}|)
	\right. \\
	&\left. \quad \quad
	+{\rm sgn}(\xi_{n\bm k}+\xi_{n'\bm k'})
	\left[
	f(-\xi_{n\bm k})-f(\xi_{n'\bm k'})
	\right]
	L_{n\bm k n'\bm k'}(|\xi_{n\bm k}+\xi_{n'\bm k'}|)
	\right]
\end{split}
\end{equation}
\end{widetext}
where $f(\xi)$ is the Fermi-Dirac distribution function, and $L$ is defined as
\begin{equation}
	\label{eq:Lfunc}
	L_{n\bm k n' \bm k'}(x) = 
	\frac{4}{\beta}\sum_{\nu>0}
	\left[
	W_{n\bm k n' \bm k'}({\rm i}\nu)
	-
	W_{n\bm k n' \bm k'}(0)
	\right]	
	\frac{x}{x^2+\nu^2}.
\end{equation}
The summation of $\nu$ in Eq. (\ref{eq:Lfunc}) is numerically evaluated following the 
procedure described in Ref.\onlinecite{RA2015}.
%It should be noted that by employing the frequency-dependent screened Coulomb interaction,
%we can evaluate the effect of static Coulomb repulsion which suppress the pairing and the 
%plasmon-driven superconducting pairing\cite{Takada1978}.

In the same way, the effect of spin fluctuations on the kernel $\mathcal K$ is decomposed as

\begin{equation}
	\begin{split}
	\label{eq:KSFdecomp}
  \mathcal K^{\rm SF}_{n\bm k n'\bm k'} &= \mathcal K^{\rm SF, stat}_{n\bm k n' \bm k'} + 
  \mathcal K^{\rm SF, dyn}_{n\bm k n'\bm k'}, \\
  \mathcal K^{\rm SF, stat}_{n\bm k n'\bm k'} &= \Lambda^{\rm SF}_{n\bm k n' \bm k'}({\rm i}\nu=0), \\
  \mathcal K^{\rm SF, dyn}_{n\bm k n'\bm k'} &= \lim_{ \{\Delta_{n\bm k}\}\to 0}
	\frac{1}{\tanh[(\beta/2)E_{n\bm k}]}\frac{1}{\tanh[(\beta/2)E_{n'\bm k'}]}
	\frac{1}{\beta^2}\\
	&\quad \times \sum_{\omega_1\omega_2}F_{n\bm k}( {\rm i}\omega_1)F_{n'\bm k'}( {\rm i}\omega_2)\\
	&\quad \quad \times\{\Lambda^{\rm SF}_{n\bm k n' \bm k'}[{\rm i}(\omega_1-\omega_2)] - \Lambda^{\rm SF}_{n\bm k n' \bm k'}({\rm i}\nu=0)\},
\end{split}
\end{equation}

where $\Lambda^{\rm SF}_{n\bm k n' \bm k'}$ is defined in the similar way as Eq. (\ref{eq:dynWnknk})
by using the effective interaction driven by spin fluctuations~\cite{Essenberger2014}
\begin{equation}
	\begin{split}
	\label{eq:LambdaSF}	
	&\Lambda^{\rm SF}(\bm r, \bm r', {\rm i}\nu) =
	-3I_{\rm xc}(\bm r,\bm r')\\
	&\quad
	-3\iint d^3r_1 d^3r_2 I_{\rm xc}(\bm r,\bm r_1)\chi_{S}(\bm r_1,\bm r_2, {\rm i}\nu)
	I_{\rm xc}(\bm r_2,\bm r').
\end{split}
\end{equation}
Here $\chi_{S}$ is the spin susceptibility obtained from the following equations ($m$ is the
spin density)
\begin{equation}
	\begin{split}
	\label{eq:chis}
	&\chi_{S}(\bm r,\bm r',{\rm i}\nu) = \chi_0(\bm r,\bm r',{\rm i}\nu) \\
	&\quad + 
	\iint d^3r_1 d^3r_2 \chi_0(\bm r,\bm r_1,{\rm i}\nu)I_{\rm xc}(\bm r_1,\bm r_2)
	\chi_{S}(\bm r_2,\bm r', {\rm i}\nu),
\end{split}
\end{equation}
and $I_{\rm xc}$ is the functional derivative of exchange-corretion functional with respect to the spin density $m$. In practice, we apply ALDA to $I_{\rm xc}$ and as a result it becomes frequency-indenpendent:
\begin{equation}
	I_{\rm xc}(\bm r, \bm r') = \frac{\delta^{2}E_{\rm xc}}{\delta m(\bm r)\delta m(\bm r')}.
	\label{Ixc}
\end{equation}
As in the Coulomb kernel,  the summation with respect to $\omega_1$ in Eq.~(\ref{eq:KSFdecomp}) is analytically calculated and we obtain
\begin{widetext}
\begin{equation}
	\begin{split}
    \mathcal K^{\rm SF, dyn}_{n\bm k n'\bm k'} &= 
	\frac{1}{\tanh[(\beta/2)\xi_{n\bm k}]}\frac{1}{\tanh[(\beta/2)\xi_{n'\bm k'}]}
	\left[
	{\rm sgn}(\xi_{n\bm k}-\xi_{n'\bm k'})
	\left[
	f(\xi_{n\bm k})-f(\xi_{n'\bm k'})
	\right]
	L^{\rm SF}_{n\bm k n'\bm k'}(|\xi_{n\bm k}-\xi_{n'\bm k'}|)
	\right. \\
	&\left. \quad \quad
	+{\rm sgn}(\xi_{n\bm k}+\xi_{n'\bm k'})
	\left[
	f(-\xi_{n\bm k})-f(\xi_{n'\bm k'})
	\right]
	L^{\rm SF}_{n\bm k n'\bm k'}(|\xi_{n\bm k}+\xi_{n'\bm k'}|)
	\right],
\end{split}
\end{equation}
\end{widetext}
where $L^{\rm SF}$ is defined by
\begin{equation}
	L^{\rm SF}_{n\bm k n' \bm k'}(x) = 
	\frac{4}{\beta}\sum_{\nu>0}
	\left[
	\Lambda^{\rm SF}_{n\bm k n' \bm k'}({\rm i}\nu)
	-
	\Lambda^{\rm SF}_{n\bm k n' \bm k'}(0)
	\right]	
	\frac{x}{x^2+\nu^2}.
\end{equation}

In the previous study, Essenberger \textit{et al.}~\cite{Essenberger2016} dropped the linear $I_{\rm xc}$ term in Eq.~(\ref{eq:LambdaSF}) on the grounds that the spin susceptibility $\chi_{S}$ is dominant in the case of Fe-based superconductors, and the first term is negligible. On the other hand, in the case of elemental metals, such an assumption is not valid and the linear $I_{\rm xc}$ term cannot be dropped. However, this term leads to physically unreasonable result as long as we use ALDA to $I_{\rm xc}$. We discuss this in detail below analogous to the screened Coulomb interaction.
In the case of the frequency-dependent screened interaction, in the high-frequency limit, 
the screened Coulomb interaction reduces to the bare Coulomb interaction because
the polarization function reduces to zero as (see Eq.~(\ref{eq:chi0})) 
\begin{align*}
	\lim_{\omega\rightarrow \infty} \chi(\textbf{r}, \textbf{r}', {\rm i}\nu\rightarrow\omega+i\delta) = 0.
\end{align*}
This frequency-dependence reflects the fact that electrons behave as bare ones: they do not feel any exchange-correlation effect when propagating with high frequencies.
According to this consideration, the effective interaction mediated by spin fluctuations originating from the exchange-correlation effect should vanish in the high-frequency limit as
\begin{align*}
	\lim_{\omega\rightarrow\infty} \Lambda^{\rm SF}(\textbf{r}, \textbf{r}', {\rm i}\nu\rightarrow\omega+i\delta)= 0.
\end{align*}
Within ALDA, however, the above condition is violated since the linear $I_{\rm xc}$ term in the left-hand side becomes $\omega$-independent and remains finite.  To avoid this problem, we neglect the first $I_{\rm xc}$ term in Eq.~(\ref{eq:LambdaSF}) in our calculations as the previous study.

Finally, the effect of spin fluctuations on the kernel $\mathcal Z$ is written as
\begin{widetext}
\begin{equation}
\begin{split}
		\mathcal Z^{\rm SF}_{n\bm k} &= -\frac{2}{\beta^2}\frac{1}{\tanh[(\beta/2)\xi_{n\bm k}]}
		\sum_{\omega_n,\omega_m}\sum_{n'\bm k'} \frac{1}{ {\rm i}\omega_n-\xi_{n\bm k}}
		\frac{1}{ {\rm i}\omega_m - \xi_{n'\bm k'}}
		\left(\frac{1}{{\rm i}\omega_n + \xi_{n\bm k}} - \frac{1}{ {\rm i}\omega_n - \xi_{n\bm k}} \right)
	\Lambda^{\rm SF}_{n\bm k n' \bm k'}[{\rm i}(\omega_n - \omega_m)] \\
		&\quad + \frac{2}{\beta^2}
		\left( \frac{1}{\xi_{n\bm k}} - \frac{\beta/2}{\sinh[(\beta/2)\xi_{n\bm k}]\cosh[(\beta/2)\xi_{n\bm k}]} \right)
	\frac{1}
	{\displaystyle \sum_{n_1 \bm k_1}\frac{\beta/2}{\cosh^2\left[(\beta/2)\xi_{n_1 \bm k_1}\right]}}
	\sum_{\omega_n \omega_m}\sum_{k'k''} \frac{1}{({\rm i}\omega_n - \xi_{n ' \bm k'})^2} 
	\frac{\Lambda^{\rm SF}_{n'\bm k' n'' \bm k''}[{\rm i}(\omega_n - \omega_m)]}{ {\rm i}\omega_m - \xi_{n'' \bm k''}}.
	\label{eq:ZSFdef}
\end{split}
\end{equation}
\end{widetext}
In the above expression, the particle-hole asymmetric component with respect to $n\textbf{k}$ is subject to divergence due to unconvergence of the $n'\textbf{k}'$ summation, as seen in $\mathcal{Z}^{\rm ph}$. Therefore, we assume the particle-hole symmetry which is
valid for many superconductors and symmetrize the above expression\cite{Luders2005, Akashi2013}. 
The symmetrization is done by replacing the $\mathcal Z^{\rm SF}_{n\bm k}=\mathcal Z^{\rm SF}(\left\{ \xi_{n\bm k} \right\})$ 
by $[\mathcal Z^{\rm SF}(\left\{ \xi_{n\bm k} \right\}) + \mathcal Z^{\rm SF}(\left\{ -\xi_{n\bm k} \right\})]/2 $.
Furthermore, we carry out the Matsubara summation with respect to
$\omega_m$ by transforming the variable as $\omega_n - \omega_m = \nu$ and rewrite 
$\mathcal Z^{\rm SF}_{n \bm k}$ as 
\begin{widetext}
\begin{equation}
\begin{split}
	\mathcal Z^{\rm SF}_{n \bm k} &= 
	-\frac{\beta/8}{\sinh[(\beta/2)\xi_{n\bm k}]\cosh[(\beta/2)\xi_{n\bm k}]} \\
	&\quad \times
	\sum_{n'\bm k'} \left[ {\rm sgn}(\xi_{n\bm k}+\xi_{n'\bm k'})L^{\rm SF}_{n\bm k n' \bm k'}(|\xi_{n\bm k} + \xi_{n'\bm k'}|)
	+ {\rm sgn}(\xi_{n\bm k}-\xi_{n'\bm k'})L^{\rm SF}_{n\bm k n' \bm k'}(|\xi_{n\bm k} - \xi_{n'\bm k'}|)\right]
	 \\
	 &\quad 
	-\frac{1}{2\tanh[(\beta/2)\xi_{n\bm k}]} \\
	 &\quad 
  \times \sum_{n'\bm k'}
	[( f(\xi_{n\bm k})-f(\xi_{n'\bm k'}) )
	 T^{\rm SF}_{n\bm k n' \bm k'}(\xi_{n\bm k} - \xi_{n'\bm k'})
	 -( f(-\xi_{n\bm k})-f(\xi_{n'\bm k'}) )
   T^{\rm SF}_{n\bm k n' \bm k'}(\xi_{n\bm k} + \xi_{n'\bm k'})],
	\label{eq:ZSFafter}
\end{split}
\end{equation}
\end{widetext}
where $T^{\rm SF}$ is defined as 
%
%\begin{equation}
%	\begin{split}
%	L_{n\bm k n' \bm k'}(x) = 
%	&\frac{4}{\beta}\sum_{\nu>0}
%	\Add{
%	\left[
%	\Lambda^{\rm SF}_{n\bm k n' \bm k'}({\rm i}\nu)
%	-
%	\Lambda^{\rm SF}_{n\bm k n' \bm k'}(0)
%	\right]	
%	}
%	\frac{x}{x^2+\nu^2}
%	&\Add{+\frac{2}{\beta}\Lambda^{\rm SF}_{n\bm k n' \bm k'}%({\rm i}\nu=0)\frac{1}{x}},
%	\label{eq:LfuncSF}
%	\end{split}
%\end{equation}
\begin{equation}
	\begin{split}
	T^{\rm SF}_{n\bm k n' \bm k'}(x) =
	& \frac{4}{\beta}\sum_{\nu>0}
	\left[
	\Lambda^{\rm SF}_{n\bm k n' \bm k'}({\rm i}\nu)
	-
	\Lambda^{\rm SF}_{n\bm k n' \bm k'}(0)
	\right]
	\frac{x^2-\nu^2}{(x^2+\nu^2)^2}.
%	&\Add{+\frac{2}{\beta}\Lambda^{\rm SF}_{n\bm k n' \bm k'}({\rm i}\nu=0)\frac{1}{x^2}},
	\label{eq:TfuncSF}
	\end{split}
\end{equation}
By analyzing this kernel closely, it turns out that 
the width of the first term significantly depends on the temperature. 
In fact, such contribution has appeared in $\mathcal Z^{\rm ph}_{n\bm k}$ and has been dropped artificially in Ref.~\onlinecite{Luders2005}. 
Here we carry out a similar procedure, namely, we neglect the first term and define $\mathcal Z^{\rm SF}_{n\bm k}$ as the second term of Eq.~(\ref{eq:ZSFafter}).
Moreover, we apply a zero temperature approximation for the even Matsubara summation. As a result, we can compute the summation as a simple integral.
Following this approximation, we obtain the expression as follows:
\begin{equation}
  \mathcal Z^{\rm SF}_{n \bm k}=\frac{1}{2}\sum_{n'\bm k'}
  T_{n {\bm k}n' \bm k'}(|\xi_{n\bm k}|+|\xi_{n'\bm k'}|).
  \label{eq:ZSF_sym_step}
\end{equation}

\section{COMPUTATIONAL DETAILS}
We used the Quantum ESPRESSO package~\cite{QE,Giannozzi_2017} to calculate the Kohn-Sham energies and wave functions.
We obtained the phonon frequencies and electron-phonon coupling by applying the density functional perturbation theory (DFPT)~\cite{DFPT}.
We used the GGA-PBE exchange-correlation functional~\cite{GGAPBE} and the optimized norm-conserving
pseudopotentials with scalar relativistic correction~\cite{ONCV} provided by Schlipf-Gygi~\cite{SG15} for all materials. Numerical conditions are listed in Table~\ref{tab:condition}. We employed the optimized tetrahedron method for the Brillouin-zone integration in all calculations~\cite{Kawamura2014}. We solved the SCDFT gap equation with the random sampling scheme~\cite{RA2012}. The sampling error is about a few percent.
In order to estimate $T_{\rm c}$, we solved the partially linearized gap equation Eq. (\ref{eq:SCDFTgap})
with increasing the temperature from zero. When we get the vanishingly small $\Delta_{n\bm k}$ solution at some temperature, we regard that temperature as the superconducting transition
temperature $T_{\rm c}$.

\begin{table}[htbp]
\caption{\label{tab:condition} Numerical conditions.}
\begin{ruledtabular}
	\begin{tabular}{cccc}
	& V & Nb & Al \\ \hline
	$\bm k$ grid\footnote{For structure and charge optimization.}
    & $18 \times 18 \times 18$
	& $16 \times 16 \times 16$
	& $16 \times 16 \times 16$ \\
	$\bm q$ grid\footnote{For dynamical matrices.} 
	& $9 \times 9 \times 9$
	& $8 \times 8 \times 8$
	& $8 \times 8 \times 8$ \\
	$\bm k$ grid\footnote{For phonon linewidth.} 
	& $36\times36\times36$
	& $32\times32\times32$
	& $32\times32\times32$ \\
	energy cutoff\footnote{For wavefunction.} 
	& 100 Ry
	& 90 Ry
	& 65 Ry \\
	energy cutoff\footnote{For charge density.} 
	& 400 Ry
	& 360 Ry
	& 260 Ry \\
\end{tabular}
\end{ruledtabular}
\end{table}

\section{Results and discussion}
\subsection{Superconducting transition temperature}
\begin{table}[!tb]
\caption{\label{tab:results} Calculation results. SF means spin fluctuations.
Values labeled ``static'' are calculated without plasmon and SF. There are some SCDFT $T_{\rm c}$ 
results calculated with the Thomas-Fermi (TF) approximation and RPA.
Values of $\mu^{\ast}$ are calculated with the improved McMillan-Allen-Dynes formula~\cite{McMillan, AllenDynes} 
using $T_{\rm c}$ calculated with SCDFT, $\lambda$, $\omega_{\rm ln}$, and $\bar{\omega}_2$.
We also show the calculated and experimental~\cite{Rietschel1979} Stoner factor $S$.}
\begin{ruledtabular}
	\begin{tabular}{cccc}
	& V & Nb & Al \\ \hline
		lattice constant [\AA] & 3.00 & 3.31 & 4.03 \\
		lattice constant [\AA] (expt.) & 3.02~\cite{Kuentzler1985} & 3.31~\cite{Laesser1985} & 4.04~\cite{Sumiyama1990} \\
		\hline
		$\lambda$ & 1.23 & 1.23 & 0.42 \\
		$\omega_{\rm ln}$ [K] & 237 & 157 & 291 \\
		$\bar{\omega}_2$ [K] & 258 & 186 & 329\\
		\hline
		$S$ &2.59 &1.64 & 1.30 \\
		$S$ (expt.)~\cite{Rietschel1979} &2.0 &1.6 & N.A. \\
		\hline
		$T_{\rm c}$ (static) [K] & 7.3 & 7.0 & 0.4 \\
		$T_{\rm c}$ (plasmon) [K] & 13.1 & 9.9 & 1.1 \\
		$T_{\rm c}$ (SF) [K] & 0.8 & 4.0 & 0.2 \\
		$T_{\rm c}$ (plasmon and SF) [K] & 1.8 & 6.0 & 0.6 \\
		\hline
		$T_{\rm c}$ (TF) [K]~\cite{Marques2005} & - & 9.5 & 0.8 \\
		$T_{\rm c}$ (RPA, static) [K]~\cite{RA2013} & - & - & 0.8 \\
		$T_{\rm c}$ (RPA, plasmon) [K]~\cite{RA2013} & - & -  & 1.4 \\
		$T_{\rm c}$ (expt.) [K]~\cite{Ashcroft} & 5.38 & 9.50 & 1.14 \\
		\hline
		$\mu^{\ast}$ (static) & 0.304 & 0.256 &  0.151 \\
		$\mu^{\ast}$ (plasmon) & 0.222 & 0.199 & 0.117 \\
		$\mu^{\ast}$ (plasmon and SF) & 0.419 & 0.277 & 0.138 \\
\end{tabular}
\end{ruledtabular}
\end{table}
\begin{figure}[!tb]
	\includegraphics[width=10truecm]{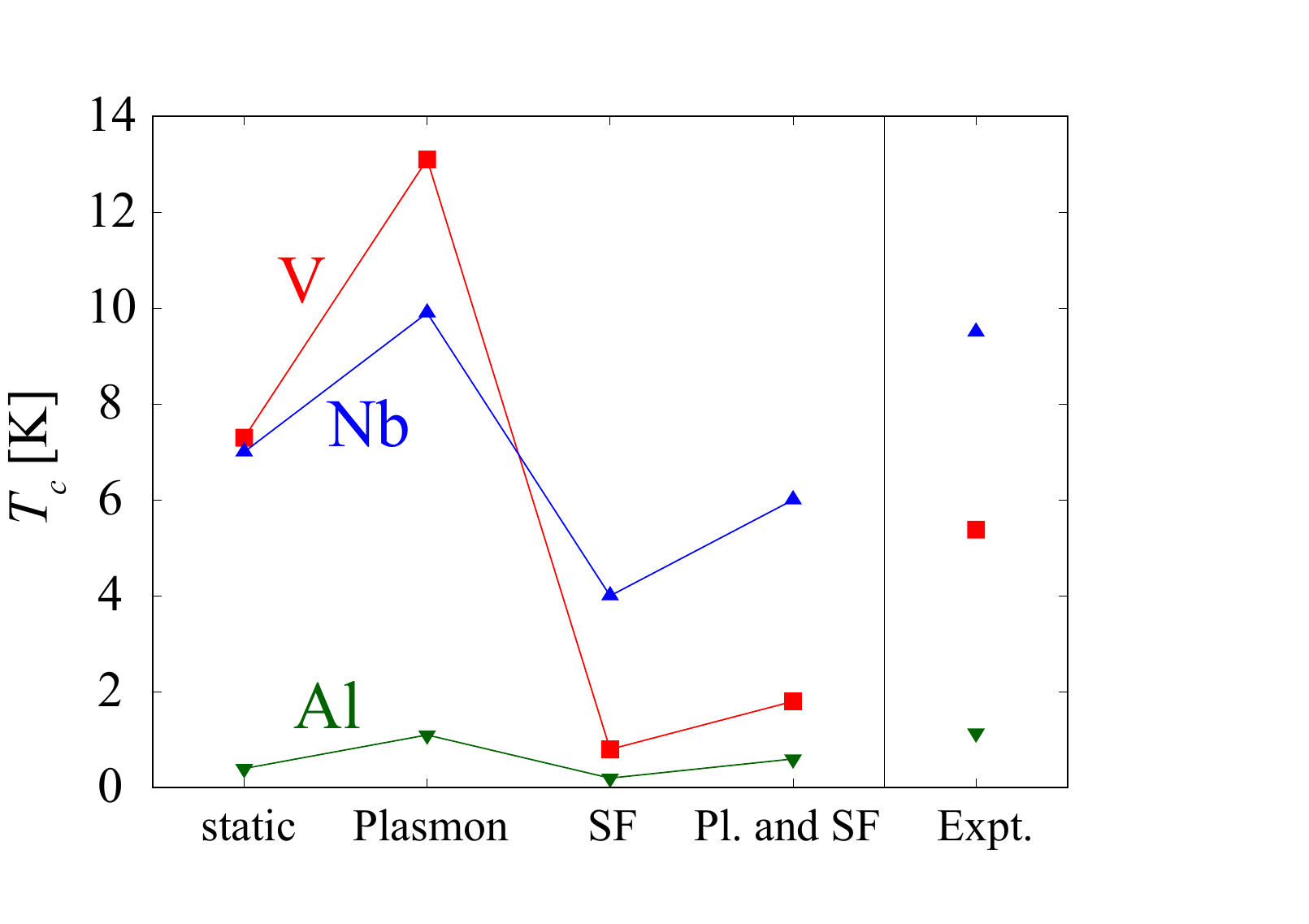}
	\caption{Comparison of calculated $T_{\rm c}$ [K] with and without the effect of plasmon 
	and spin fluctuation and experimental $T_{\rm c}$ in each material. 
Red, blue and Green points are data for V, Nb and Al, respectively.}
	\label{fig:Tcdiff}
\end{figure}
Calculated results are summarized in Table~\ref{tab:results}.
The lattice constants resulting from the structure optimization agree well with the experimental ones.
We use the theoretically optimized lattice constants in the following calculations.

In the case of V, we obtain $T_{\rm c} =$ 7.3 K without the effects of both of plasmon and spin fluctuations. This value is already higher than experimental data by about 35\%.
When we calculate $T_{\rm c}$ with the effect of plasmon, $T_{\rm c}$ is enhanced to 13.1 K, which is extremely higher than the experimental value. On the other hand, when we calculate $T_{\rm c}$ with the effect of spin fluctuations, $T_{\rm c}$ is suppressed to 0.8 K. Furthermore, when we include both of the effect of plasmon and spin fluctuations, $T_{\rm c}$ becomes 1.8 K, which is lower than experimental one by about 70 \%.
In the case of Nb, $T_{\rm c}$ is 7.0 K without plasmon and spin fluctuations.
When we include the effect of plasmon or spin fluctuations, we obtain $T_{\rm c}$ = 9.9, 4.0 K, respectively. If both of the effects are included, $T_{\rm c}$ is reduced to 6.0 K. Finally, in Al, calculated $T_{\rm c}$ changes from 0.4 K to 1.1, 0.2 K by including the effect of plasmon or spin fluctuations. When we consider both, $T_{\rm c}$ settles into $0.6$ K. The quantitative difference of $T_c$ between the present result and Ref.~\cite{Kawamura2020} is probably due to the variation in the treatment of kernels originating from electron-phonon interaction, $\mathcal{K}^{\rm ph}$ and $\mathcal{Z}^{\rm ph}$. Note that we adopted energy-averaged functionals as Eqs. (\ref{eq:Kphdef}) and (\ref{eq:Zphdef}), while fully $(n, \bm{k})$-dependent functionals in the vicinity of Fermi surface are used in Ref.~\cite{Kawamura2014}.

We summarize the data of calculated $T_{\rm c}$ in each condition and 
experimental value in Fig.~\ref{fig:Tcdiff}. This figure shows some common features among these three materials. For example, we can see that the effect of plasmon causes the enhancement of $T_{\rm c}$. This plasmon-assisted superconducting mechanism is originally studied by Takada~\cite{Takada1978},
and later studied within the SCDFT scheme by Akashi {\it et al}~\cite{RA2013}. The present results are consistent
with the previous studies. Another important feature is that the effect of spin fluctuations causes the reduction of $T_{\rm c}$. This behavior is also consistent with the previous study by Berk and Schrieffer~\cite{Berk1966}. Although this qualitative trend is common, the amount of reduction differs among materials. Specifically, we can see a relatively strong reduction in V.
This quantitatively different behavior is interpreted to be derived from the degree of localization of the metallic electron orbitals.

The ferromagnetic fluctuations originate from the Pauli exchange effect~\cite{Berk1966}: 
Two electrons having the spins in the same direction are required to be separated in space due to the Pauli principle. As a result, in a nearly homogeneous system, these two electrons have lower energy than that of two electrons having opposite spins, and the ferromagnetic fluctuations are induced.
In addition, the exchange effect is significant in spatially localized electronic states.
Returning to the present results, the valence states are dominated by 4$d$ states in Nb, 
and by 3$d$ in V, respectively. 
In the latter case, the valence states are more localized.
Therefore, the effect of spin fluctuations should be stronger in V than in Nb.
The result in Al can be interpreted similarly. In the case of Al, the $T_{\rm c}$ reduction due to spin fluctuations
is less appreciable. This result is reasonable because the valence states are formed by less localized
3$s$ and 3$p$ states in Al, and the exchange effect is relatively smaller than in the case of V and Nb.

It should be emphasized that we were able to reproduce the experimental fact that the $T_{\rm c}$ of V is considerably lower than that of Nb only by considering the effect of spin fluctuations. This result suggests that spin fluctuations play an essential role in determining $T_{\rm c}$ of transition metals. On the other hand, the calculated values of $T_{\rm c}$ with the effect of spin fluctuations are significantly underestimated compared to the experimental values in V, Nb. We can trace back the reason for this underestimation to the approximation used for the exchange-correlation kernel.

In Table~\ref{tab:results}, we present the calculated and experimental~\cite{Rietschel1979} values of Stoner enhancement factor $S$, which is the good measure for the strength of ferromagnetic fluctuations:
\begin{equation}
	S = \frac{\chi_S(\bm q =0, \bm q' =0, {\rm i}\nu =0)}{\chi_0(\bm q =0, \bm q' =0, {\rm i}\nu =0)}.
\end{equation}
We can see the calculated values are larger than the experimental ones in V and Nb. This means the spin susceptibility $\chi_S$ and therefore the effect of spin fluctuations are overestimated in the present calculation. Since the accuracy of $\chi_S$ in Eq. (\ref{eq:chis})  is determined by the approximation to the exchange-correlation kernel $I_{\rm xc}$, the origin of the discrepancy is ascribed to the ALDA to $I_{\rm xc}$ in Eq. (\ref{Ixc}). The disagreement could be solved by using more accurate functional, but that is out of the scope of this paper and left as future work.

We also calculate the values of $\mu^{\ast}$ based on the improved McMillan-Allen-Dynes formula~\cite{AllenDynes}
using calculated $\lambda, \omega_{\rm ln}, \bar{\omega}_2$ listed in Table~\ref{tab:results} to reproduce 
the $T_{\rm c}$ calculated in SCDFT. 
The calculated values of $\mu^{\ast}$ with the effect of spin fluctuations for V and Nb are 
0.419 and 0.277, respectively. In the analysis with ME theory, this value is conventionally
set to about 0.10~\cite{Carbotte1990}. Thus, the conventional value does not apply to those systems because of the spin-fluctuations effect.

\subsection{Analysis of the kernels}
\subsubsection{$\mathcal{K}$ : off-diagonal part of kernel}
We thus find that the effect of spin fluctuations on $T_{\rm c}$ is at least qualitatively reasonable and confirm that the amount of the $T_{\rm c}$ reduction reflects the magnitude of localization of electron orbitals.
In order to investigate how spin fluctuations affect the $T_{\rm c}$ closely, we 
consider the partially averaged nondiagonal exchange-correlation kernel, which is defined as
\begin{equation}
	\mathcal K_{n\bm k}(\xi) \equiv \frac{1}{N(\xi)}
	\sum_{n'\bm k'} \mathcal K_{n\bm k n'\bm k'}\delta(\xi - \xi_{n'\bm k'}).
	\label{eq:aveK}
\end{equation}

\begin{figure}[!tb]
	\centering
	\includegraphics[width=0.45\textwidth,clip]{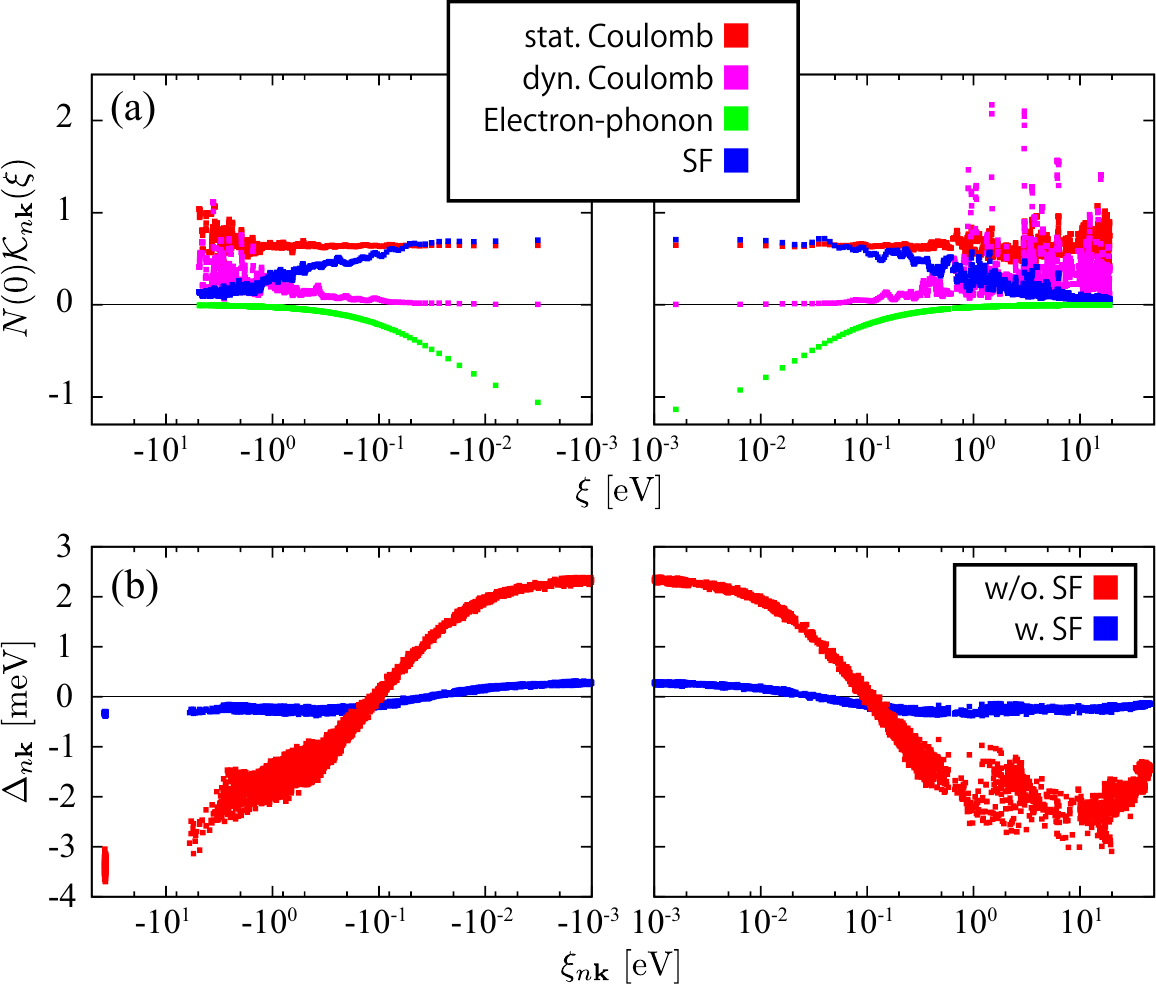}
	\caption{(a) Decomposition of the averaged exchange-correlation kernels defined in 
		Eq.~(\ref{eq:aveK}) at $T=0.01$K in V. An arrow indicates the peak in the kernel
		originating from spin fluctuations.
		(b) Corresponding gap functions without 
	and with SF. $\xi$ is the energy measured from the chemical potential $\mu$. }
	\label{fig:Vaveker}
\end{figure}

The averaged exchange-correlation kernels and corresponding gap functions calculated for V 
are plotted in Fig.~\ref{fig:Vaveker}. The kernel originating from the electron-phonon interaction
$\mathcal K^{\rm ph}$ is negative within the low energy region and nearly zero in the region where $\xi \gtrsim 1$eV. The kernel of the static Coulomb interaction is nearly constant in the
whole energy region. On the other hand, the kernel stemming from the plasmon-induced dynamical
Coulomb interaction rises from nearly zero to positive around $\xi \approx 10^{-1}$eV.
This contribution suppresses the effective repulsion between quasiparticles due to the retardation effect, and the resulting $T_{\rm c}$ is enhanced compared to that of static approximation~\cite{RA2013}.
The kernel originating from dynamical spin fluctuations is a nearly positive constant in the low-energy region. 
In the high-energy region, on the other hand, the contribution from spin fluctuations reduces to zero. 
Furthermore, the magnitude of repulsion due to spin fluctuations in the low-energy region is comparable to that of the static Coulomb interaction. This energy dependence of
$\mathcal K^{\rm SF}$ suppresses strongly the low-energy positive gap function (Fig.~\ref{fig:Vaveker}(b)) and the resulting $T_{\rm c}$.

%
%\begin{figure}[h]
%	\centering
%	\includegraphics[width=10truecm,clip]{V_gap_mod.eps}
%	\caption{Temperature dependence of the gap functions calculated with and without 
%		the spin fluctuations kernel $\mathcal K^{\rm SF}$ and the experimental $T_c$ in V.
%		An arrow indicates the experimental $T_c = 5.38{\rm K}$.}
%	\label{fig:VgapT}
%\end{figure}
%

\begin{figure}[!tb]
	\centering
	\includegraphics[width=0.45\textwidth,clip]{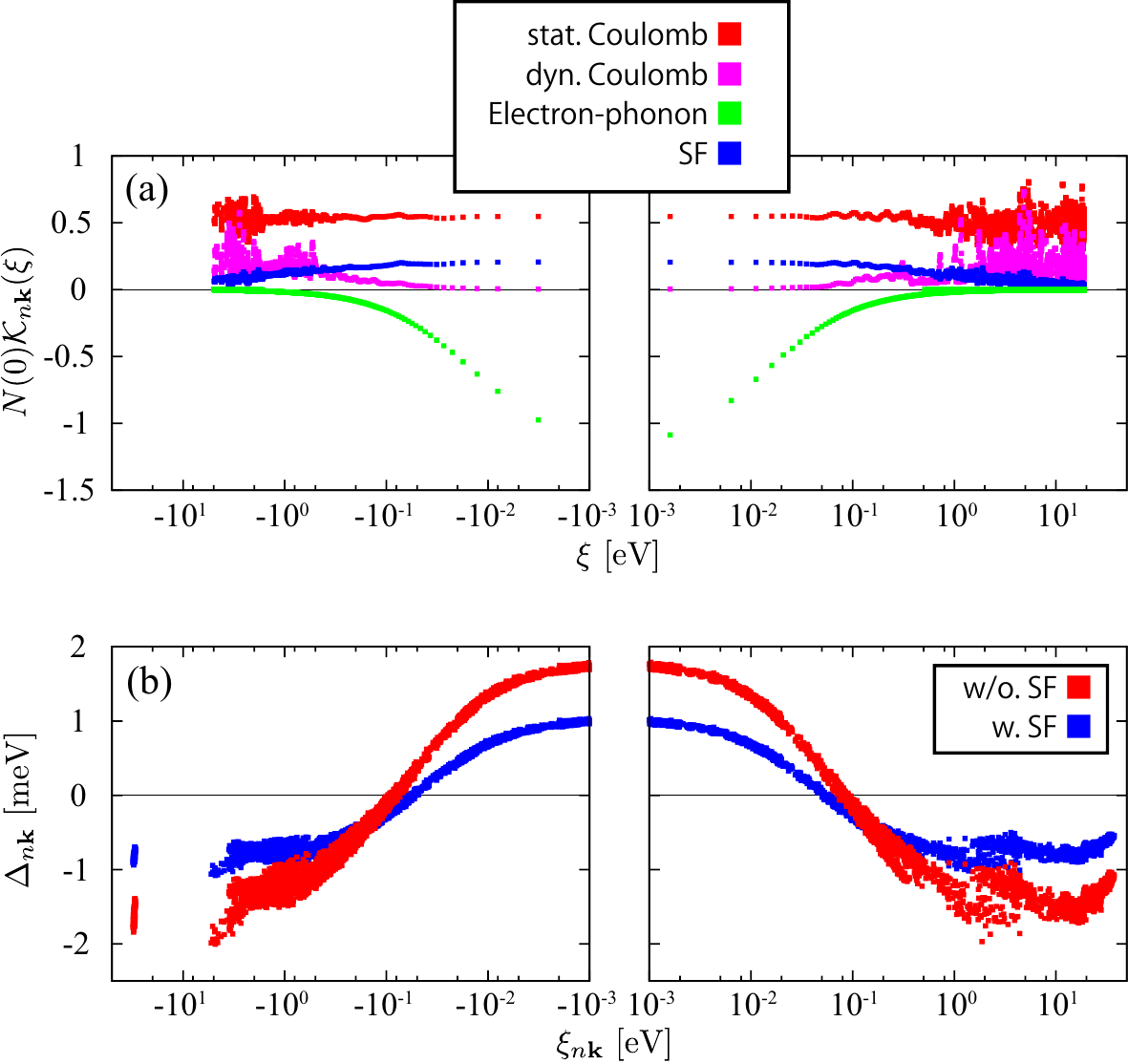}
	\caption{(a) Decomposition of the averaged exchange-correlation kernels
		at $T=0.01$K in Nb. (b) Corresponding gap funtions without 
	and with SF.}
	\label{fig:Nbaveker}
\end{figure}

Next, we see the averaged exchange-correlation kernels calculated for Nb in Fig.~\ref{fig:Nbaveker}.
In the case of Nb, the qualitative trend is the same as in the case of V: The averaged kernel of spin fluctuations is a nearly positive constant in low energy region and reduces to zero in high energy region.
However, the magnitude of the kernel $\mathcal K^{\rm SF}$ is relatively smaller than that of V.
As noted above, this is due to the strength of the electronic localization.
%
%\begin{figure}[h]
%	\centering
%	\includegraphics[width=10truecm,clip]{Nb_gap_mod.eps}
%	\caption{Temperature dependence of the gap functions calculated with and without 
%		the spin fluctuations kernel $\mathcal K^{\rm SF}$ and the experimental $T_c$ in Nb.
%		An arrow indicates the experimental $T_c = 9.5{\rm K}$.}
%	\label{fig:NbgapT}
%\end{figure}

\begin{figure}[!tb]
	\centering
	\includegraphics[width=0.45\textwidth,clip]{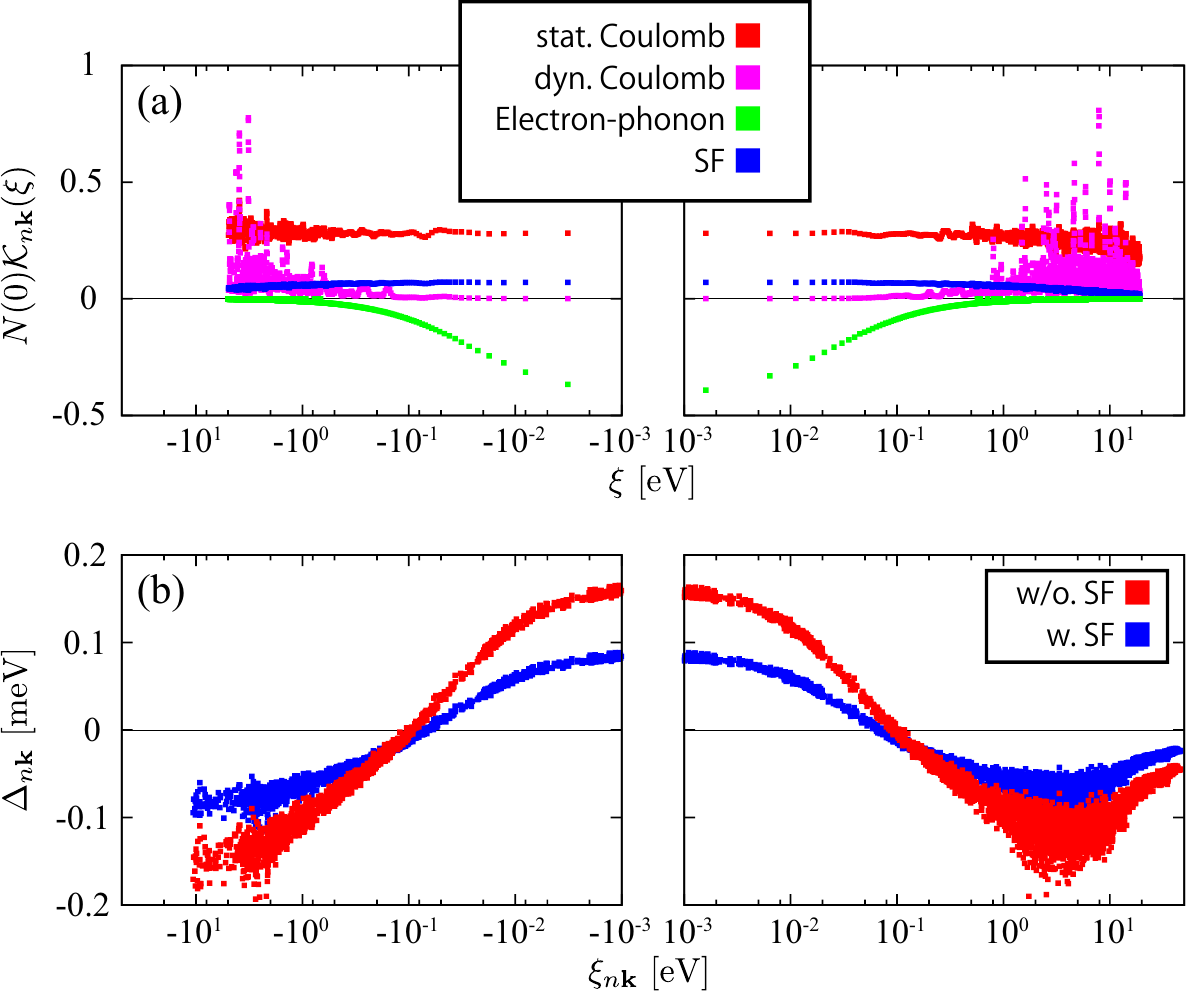}
	\caption{(a) Decomposition of the averaged exchange-correlation kernels
		at $T=0.01$K in Al. (b) Corresponding gap funtions without 
	and with SF.}
	\label{fig:Alaveker}
\end{figure}
Finally, we see the averaged exchange-correlation kernels calculated for Al in Fig.~\ref{fig:Alaveker}.
The kernel $\mathcal K^{\rm SF}$ is significantly smaller than that of the static Coulomb interaction.
%
% kokokara DISCUSSION deha ?????
%

%
%\begin{figure*}
%	\includegraphics[width=12truecm]{V_kernels.pdf}% Here is how to import EPS art
%	\caption{\label{Vaveker}(a) Decomposition of the averaged exchange-correlation kernels defined in 
%		Eq.(\ref{eq:aveK}) at $T=0.01$K in V. An arrow indicates the peak in the kernel
%		originating from spin fluctuations.
%		(b) Corresponding gap funtions without 
%	and with SF. $\xi$ is the energy measured from the chemical potential $\mu$. An arrow
%indicates the cusp in the gap function.}
%\end{figure*}
%

\begin{table}
\begin{tabular}{l | c c c}
\hline
\hline
& V & Nb & Al \\ \hline
$\mathcal{Z}^{\rm SF}(0)$ &0.177&$4.68\times10^{-2}$&$2.40\times 10^{-2}$ \\
$N(0)\mathcal{K}^{\rm SF}(0,0)$ &0.699&0.198&$6.92\times 10^{-2}$ \\ \hline\hline
\end{tabular}
\caption{The calculated values of averaged diagonal kernel $\mathcal{Z^{\rm SF}}$ and nondiagonal kernel $\mathcal{K^{\rm SF}}$ at the Fermi surface.}
\label{tab:kernels-SF}
\end{table}

\begin{figure}[!tb]
	\begin{center}
	\includegraphics[width=8.5truecm]{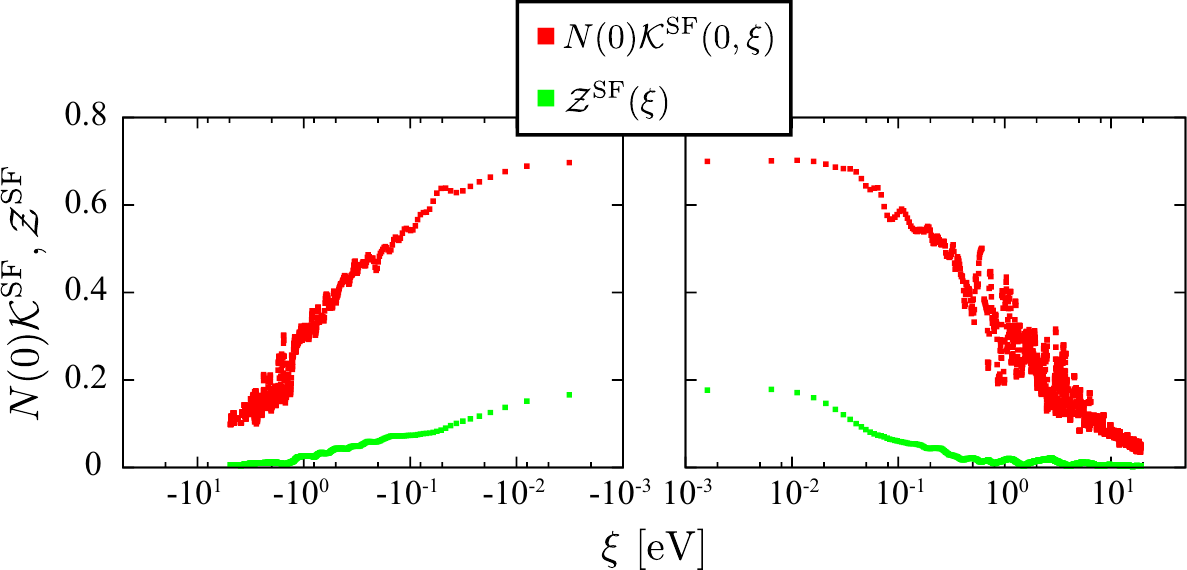}
	\caption{Comparison between the energy dependence of averaged diagonal kernel $\mathcal{Z}^{\rm SF}$ and nondiagonal kernel $\mathcal{K}^{\rm SF}$ in the case of V.}
	\label{fig:Zsf-NKsf-v}
	\end{center}
\end{figure}

\subsubsection{$\mathcal{Z}$ : diagonal part of kernel}
We next analyze the mass-renormalization part $\mathcal{Z}$ for the SF part. For the phonon-mediated interaction, it is well known that the pairing strength $\lambda$ and mass-renormalization of normal electronic states $m \rightarrow m^{\ast}$ are closely related by the formula $m^{\ast}\simeq m(1+\lambda)$~\cite{McMillan}. In SCDFT, this relation is represented by $-N(0)\mathcal{K}^{\rm ph}(0,0)\simeq \mathcal{Z}^{\rm ph}(0)\simeq \lambda$, where $\mathcal{K}(\xi, \xi')$ and $\mathcal{Z}(\xi)$ are the averaged kernels defined as
\begin{align}
\nonumber
\mathcal{K}^{\rm SF}(\xi,\xi') 
&\!\equiv\!
\frac{1}{N(\xi)N(\xi')} \\
&\times\sum_{n{\bm k}n'{\bm k}'}\!\!
\delta(\xi \!-\!\xi_{n{\bm k}})\delta(\xi' \!\!-\!\xi_{n'{\bm k}'})\mathcal{K}^{\rm SF}_{n{\bm k}n'{\bm k}'},
\\
\mathcal{Z}^{\rm SF}(\xi) 
&\!\equiv\!
\frac{1}{N(\xi)} \!\!
\sum_{n{\bm k}}\!
\delta(\xi \!-\!\xi_{n{\bm k}})\mathcal{Z}^{\rm SF}_{n{\bm k}}
.
\end{align}
This condition is automatically satisfied by the approximation to the electron-phonon coupling in Eq.~(\ref{eq:Eliashbergfunc})~\cite{Luders2005}. In contrast, we have not employed any approximations to the momentum and band dependence on $\Lambda^{\rm SF}_{n{\bf k}n'{\bf k}'}({\rm i}\nu)$. We find that the explicit treatment of this dependence yields a relation between the SF-mediated pairing part $N(0)\mathcal{K}^{\rm SF}(0,0)$ and the mass renormalization part $\mathcal{Z}^{\rm SF}(0)$ different from the phonon case.

The calculated values of averaged kernels $N(0)\mathcal{K}^{\rm SF}(0,0)$ and $\mathcal{Z}^{\rm SF}(0)$ are summarized in Table~\ref{tab:kernels-SF}, and their energy dependences especially in the case of V are depicted in Fig.~\ref{fig:Zsf-NKsf-v}. Generally, $\mathcal{Z}^{\rm SF}(0)$ is systematically smaller than $N(0)\mathcal{K}^{\rm SF}(0,0)$ by factor of $1/3$--$1/4$. In order to clarify this origin, we analyze these averaged kernels below.

We here define the averaged matrix element $\tilde{\Lambda}^{\rm SF}$ by
%\begin{eqnarray}
%\sum_{n{\bm k}n'{\bm k}'}\!\!
%\delta(\xi \!-\!\xi_{n{\bm k}})\delta(\xi' \!\!-\!\xi_{n'{\bm k}'})
%\Lambda^{\rm SF}_{n{\bm k}n'{\bm k}'}(i\nu)
%\equiv
%N(\xi)N(\xi')\tilde{\Lambda}^{\rm SF}(\xi,\xi'i\nu)
%.
%\end{eqnarray}
\begin{align}
\nonumber
\tilde{\Lambda}^{\rm SF}(\xi, \xi', {\rm i}\nu)
\equiv&\frac{1}{N(\xi)N(\xi')}\\
&\times\sum_{n{\bm k}n'{\bm k}'}\!\!
\delta(\xi \!-\!\xi_{n{\bm k}})\delta(\xi' \!\!-\!\xi_{n'{\bm k}'})
\Lambda^{\rm SF}_{n{\bm k}n'{\bm k}'}({\rm i}\nu).
\end{align}
The resulting forms of the averaged kernels at the zero-temperature limit are as follows:
\begin{eqnarray}
\mathcal{K}^{\rm SF}(\xi, \xi')
&=&\frac{1}{\pi}\int d\nu 
\tilde{\Lambda}^{\rm SF}(\xi, \xi', {\rm i}\nu) \frac{|\xi |+|\xi'|}{(|\xi |+|\xi'|)^2+\nu^2},
\label{eq:KSF-ave}
\\
\nonumber
\mathcal{Z}^{\rm SF}(\xi)
&=&\frac{1}{2\pi}\int d\xi' N(\xi') \int d\nu \\
\nonumber
&& \quad \times \left[ \tilde{\Lambda}^{\rm SF}(|\xi|, |\xi'|, {\rm i}\nu) - \tilde{\Lambda}^{\rm SF}(|\xi|, |\xi'|, 0) \right]\\
&& \quad \times
 \frac{(|\xi |+|\xi'|)^2 -\nu^2}{[(|\xi |+|\xi'|)^2+\nu^2]^2}
 .
 \label{eq:ZSF-ave}
\end{eqnarray}
In the latter, we inserted the identity $1=\int d\xi' \delta(\xi' - \xi_{n'{\bf k}'})$. The factor $\tilde{\Lambda}^{\rm SF}(\xi, \xi', {\rm i}\nu)$ generally decays with respect to $\xi, \xi', \nu$; in other words, it has energy cutoffs. We simply treat this aspect by the following approximation

\begin{widetext}
\begin{eqnarray}
\frac{N(\xi)}{N(0)}\frac{N(\xi')}{N(0)}\tilde{\Lambda}^{\rm SF}(\xi,\xi', {\rm i}\nu)
\simeq
\left\{
\begin{array}{l c}
\theta(\nu_{\rm cut}-|\nu|)\tilde{\Lambda}^{\rm SF}(0,0,0) & \ \ (-L_{1} \leq \xi \leq L_{2} \ \  {\rm and} \ \   -L_{1} \leq \xi' \leq L_{2}) \\
0 &({\rm otherwise})
\end{array}
\right. 
.
\label{eq:LambdaSF-approx}
\end{eqnarray}
\end{widetext}
Parameters $L_{1}, L_{2}, \nu_{\rm cut}$ are the energy cutoffs. With this approximation, in the low-temperature limit, we get the following (see Appendix~\ref{sec:derive-ZtoNK} for the detail of its derivation)
\begin{eqnarray}
\frac{\mathcal{Z}^{\rm SF}(0)}{N(0)\mathcal{K}^{\rm SF}(0,\!0)}
=
1
-\frac{1}{\pi}
\left(
{\rm arctan}\frac{\nu_{\rm cut}}{L_{1}}
\!+\!
{\rm arctan}\frac{\nu_{\rm cut}}{L_{2}}
\right)
.
\label{eq:ZtoNK-ratio}
\end{eqnarray}
This equation indicates that the mass-renormalization is rendered ``incomplete" when the decay rates of $\tilde{\Lambda}^{\rm SF}(\xi, \xi', {\rm i\nu})$ with respect to the respective energy variables are competing. 

\begin{figure}[!tb]
	\centering
	\includegraphics[width=9.5truecm,clip]{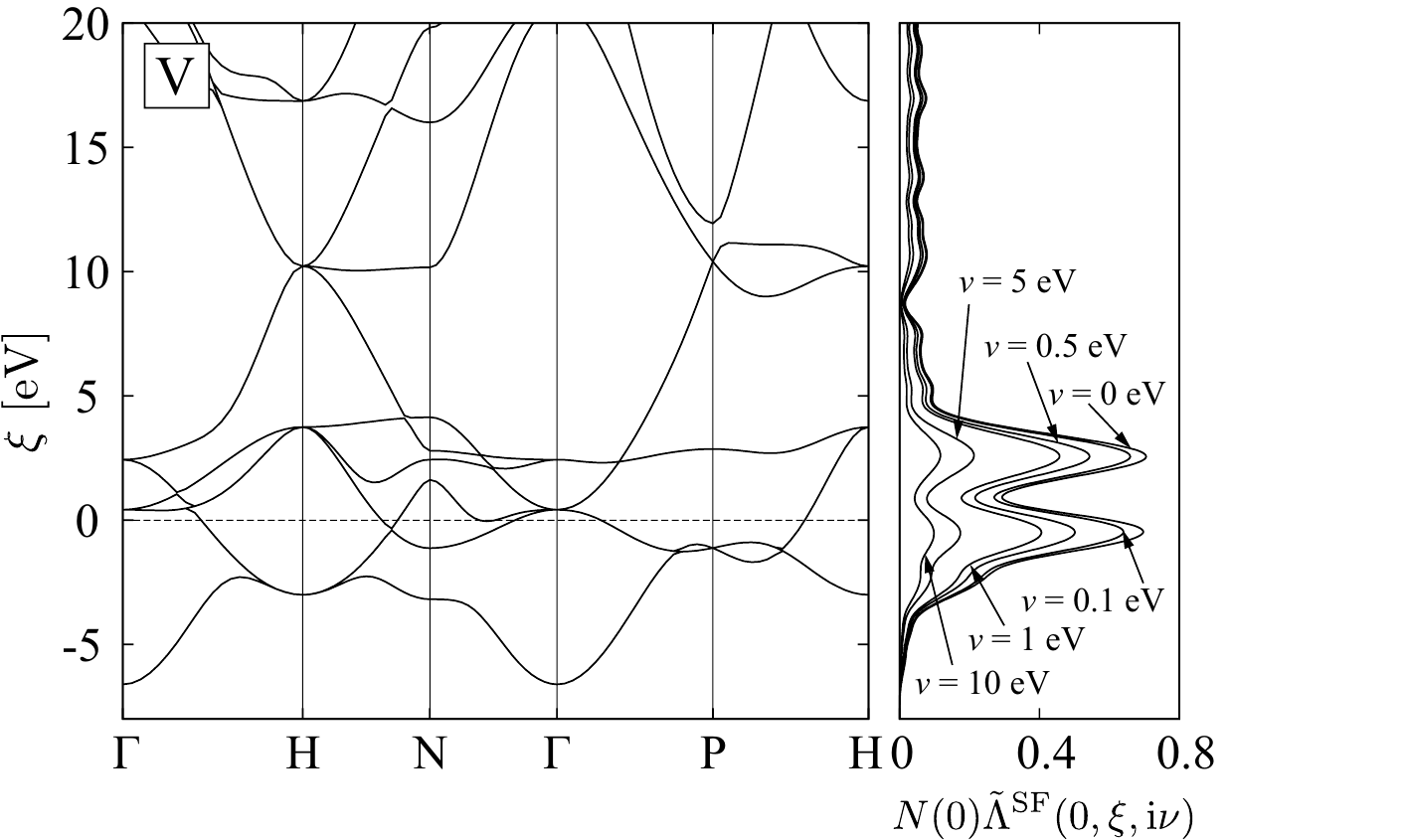}
	\caption{(Left) Band structure of V.
	  (Right) Averaged $\Lambda^{\rm SF}$ times the density of states
      at the Fermi level.}
	\label{fig:band_v}
\end{figure}

\begin{figure}[!tb]
	\centering
\includegraphics[width=9.5truecm, clip]{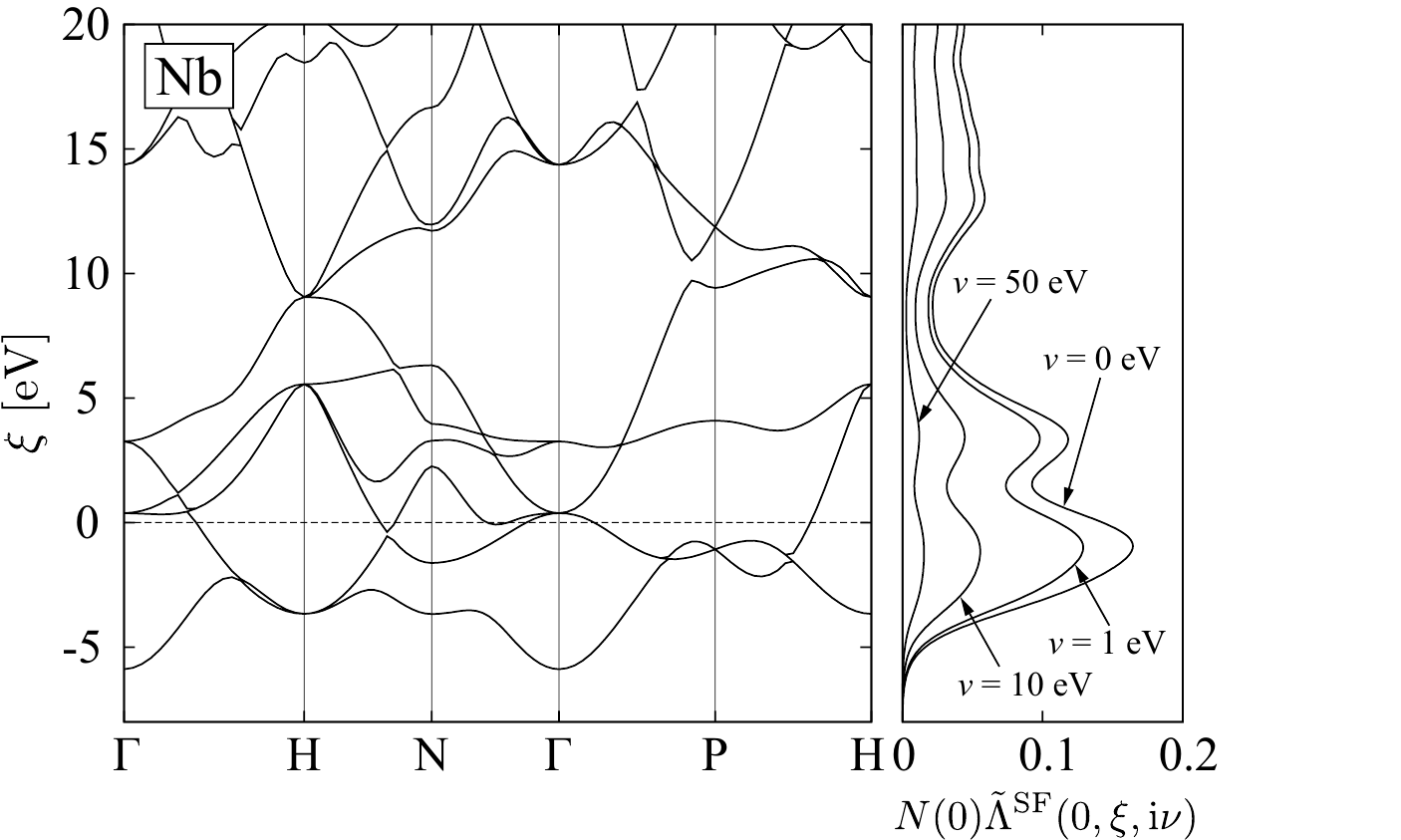}
	\caption{(Left) Band structure of Nb.
	  (Right) Averaged $\Lambda^{\rm SF}$ times the density of states
      at the Fermi level.}
	\label{fig:band_nb}
\end{figure}

\begin{figure}[!tb]
	\centering
	\includegraphics[width=9.5truecm,clip]{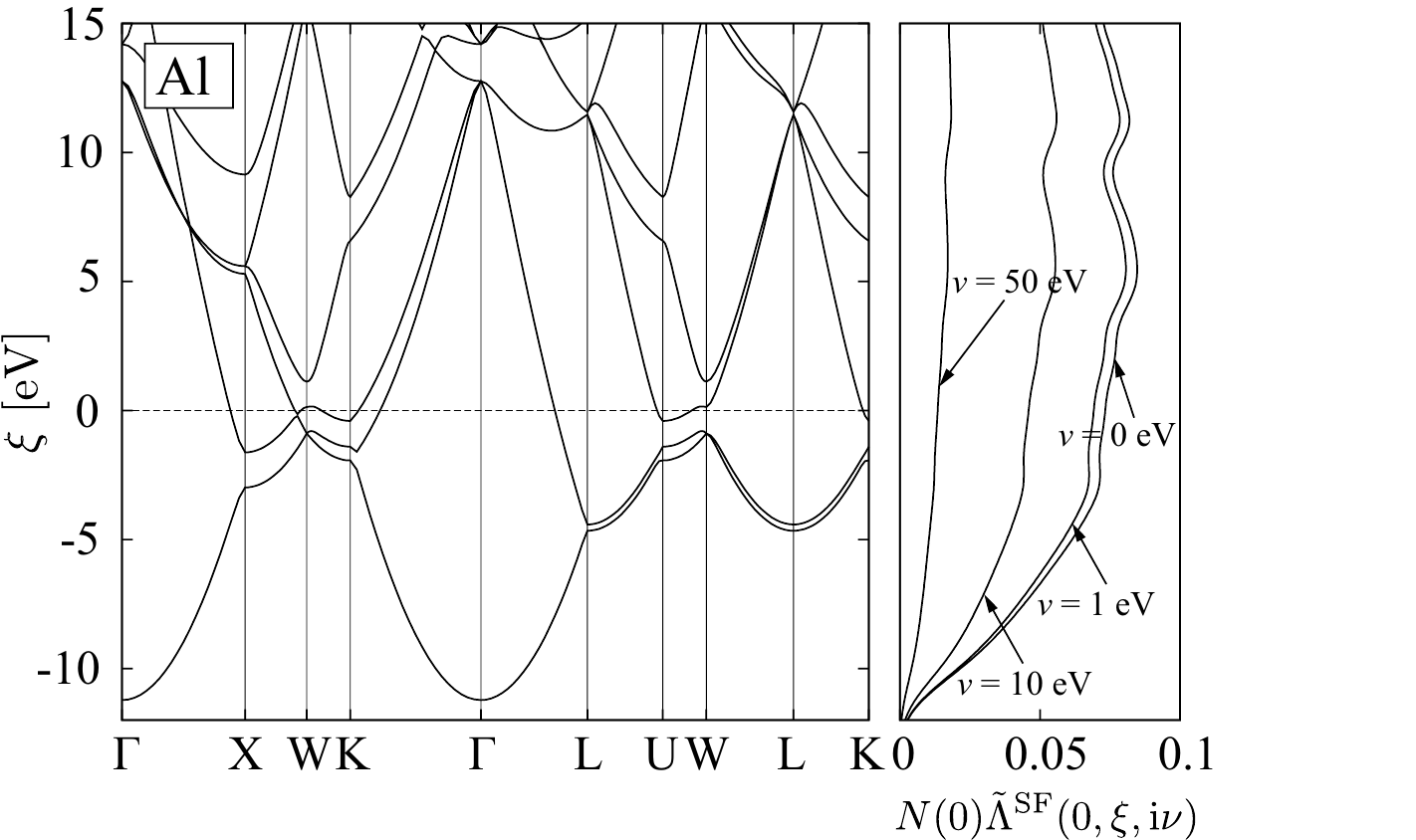}
	\caption{(Left) Band structure of Al.
	  (Right) Averaged $\Lambda^{\rm SF}$ times the density of states
      at the Fermi level.}
	\label{fig:band_al}
\end{figure}

$\tilde{\Lambda}^{\rm SF}(\xi=0, \xi', {\rm i}\nu)$ with various $\nu$ for V, Nb and Al are plotted in Figs.~\ref{fig:band_v}, \ref{fig:band_nb}, \ref{fig:band_al} together with the band structure of these materials. Commonly, this function shows slowly decaying behavior with respect to $\nu$, where its full width at half maximum is roughly estimated to be a few tens of eV. This is the similar energy scale to the decaying length with respect to $\xi'$. Especially, in V and Nb, $\tilde{\Lambda}^{\rm SF}(\xi=0, \xi', {\rm i}\nu)$ has large values  where $\xi'$ is in the energy range of the bands originating from the $3d$ and $4d$ orbitals. It is thus confirmed that the difference between the calculated values of $\mathcal{Z}^{\rm SF}(0)$ and $N(0)\mathcal{K}^{\rm SF}(0,0)$ originates from the competition of the energy scales dominating $\tilde{\Lambda}^{\rm SF}$. This is primarily due to the fact that the frequency of the paramagnetic collective modes (paramagnon) has the energy scale of normal electrons, different from the phonon kernels where the decaying length with respect to $\nu$ is of order of the phonon energy scale. We therefore infer that the ``incomplete" mass-renormalization for the SF part generally occurs and then the conventional averaging approximation~\cite{Essenberger2014},
%\begin{eqnarray}
%\sum_{n{\bm k}n'{\bm k}'}\!\!
%\delta(\xi \!-\!\xi_{n{\bm k}})\delta(\xi' \!\!-\!\xi_{n'{\bm k}'})
%\Lambda^{\rm SF}_{n{\bm k}n'{\bm k}'}(i\nu)
%\equiv
%N(0)N(0)\tilde{\Lambda}^{\rm SF}(i\nu)
%,
%\end{eqnarray}
\begin{align}
\nonumber
\tilde{\Lambda}^{\rm SF}_{\rm Ave}({\rm i}\nu)
&\equiv
\frac{1}{N(0)N(0)}
\sum_{n{\bm k}n'{\bm k}'}\!\!
\delta(\xi_{n{\bm k}})\delta(\xi_{n'{\bm k}'})
\Lambda^{\rm SF}_{n{\bm k}n'{\bm k}'}({\rm i}\nu)
,
\end{align}
yields overestimation of $\mathcal{Z}^{\rm SF}$.

\section{Summary}

In this work, we have studied the spin-fluctuation effect on $T_{\rm c}$ of conventional superconductors V, Nb, and Al with extended density functional theory for superconductors. The calculated values of $T_{\rm c}$ have shown that ferromagnetic fluctuations compete with superconductivity induced by the cooperation between phonons and plasmons. In particular, we have confirmed that the amount of reduction of $T_{\rm c}$ due to spin fluctuations reflects the degree of localization of valence states. Furthermore, we have shown that the inclusion of the spin-fluctuations effect enables us to reproduce the relation between $T_{\rm c}$'s of V and Nb successfully, while it leads to the significant underestimation of $T_{\rm c}$.  This seems to be caused by the overestimation of spin susceptibility within the current DFT scheme. Finally, we have analyzed the exchange-correlation kernels, and it has been revealed that the mass-renormalization of spin fluctuations is incomplete, which is different from that of phonons. Our results establish the quantitative effect of spin fluctuations on elemental superconductors, which base the future developments of first-principles calculation of $T_{\rm c}$.

Note: When we were finalizing this manuscript we became aware of a study\cite{Sanna2020}, where an improved form of $\mathcal{K}^{\rm ph}, \mathcal{Z}^{\rm ph}$  kernels have been proposed. Use of this may improve the agreement of the calculated and experimentally observed $T_{\rm c}$'s.

\begin{acknowledgments}
The numerical calculations in this paper were performed on the supercomputers in ISSP and the Information Technology Center at the University of Tokyo.  Y.H. was supported by the Japan Society for the Promotion of Science through the Program for Leading Graduate Schools (MERIT). This work was partly supported by MEXT Elements Strategy Initiative to Form Core Research Center (grant no. JPMXP0112101001). This work was supported by JSPS KAKENHI Grant Numbers 15K20940 and 20K15012 from Japan Society for the Promotion of Science (JSPS).
\end{acknowledgments}

\appendix

\section{Derivation of Eq.~(\ref{eq:ZtoNK-ratio})}
\label{sec:derive-ZtoNK}
For the expressions Eqs.~(\ref{eq:KSF-ave}) and (\ref{eq:ZSF-ave}), we here consider the limit $\xi, \xi' \rightarrow 0+$ with $|\xi|, |\xi'| \gg T$. Note that the limiting behavior of the kernels does not concern from which side the limit is taken and it is enough to consider a specific case where the energy variables are positive. Applying the approximation Eq.~(\ref{eq:LambdaSF-approx}), we first obtain
\begin{eqnarray}
\mathcal{K}^{\rm SF}(0,0)
&\equiv&
\lim_{\xi, \xi' \rightarrow 0+}
\mathcal{K}^{\rm SF}(\xi,\xi')
\nonumber \\
&=&
\frac{1}{\pi}
\tilde{\Lambda}^{\rm SF}(0,0,0)
\lim_{\xi \rightarrow 0+}
\int_{-\nu_{\rm cut}}^{\nu_{\rm cut}} d\nu
\frac{\xi}{\nu^2+\xi^2}
\nonumber \\
&=&
\tilde{\Lambda}^{\rm SF}(0,0,0)
\label{eq:KSF-analytic}
.
\end{eqnarray}
Above, we used the fact $\lim_{\xi\rightarrow 0+} {\rm arctan}(\nu_{\rm cut}/\xi)=\pi/2$. We also apply the same approximation to $\mathcal{Z}^{\rm SF}(\xi)$. In this case, it has to be noted that the cutoffs $L_{1}$ and $L_{2}$ limits the range of the integral with respect to $\xi'$. We consequently get
%\newpage
%\begin{widetext}
\begin{eqnarray}
\mathcal{Z}^{\rm SF}(0)
&\equiv &
\lim_{\xi \rightarrow 0+} \mathcal{Z}^{\rm SF}(\xi)
\nonumber\\
&=&
-\frac{N(0)\tilde{\Lambda}^{\rm SF}(0,0,0)}{\pi }
\int_{-L_{1}}^{L_{2}} d\xi'
\int_{\nu_{\rm cut}}^{\infty} d\nu
\frac{\xi'^2-\nu^2}{[\xi'^2+\nu^2]^2}.
\nonumber\\
\end{eqnarray}
The integrations with respect to $\xi'$ and $\nu$ above are analytically carried out and we get
\begin{eqnarray}
&&\lim_{\xi \rightarrow 0+} \mathcal{Z}^{\rm SF}(\xi)
\nonumber\\
&&=
\frac{N(0)\tilde{\Lambda}^{\rm SF}(0,0,0)}{\pi}
\left\{
\pi
-
{\rm arctan}\frac{\nu_{\rm cut}}{L_{1}}
-
{\rm arctan}\frac{\nu_{\rm cut}}{L_{2}}
\right\}.
\nonumber\\
\label{eq:ZSF-analytic}
\end{eqnarray}
%\end{widetext}
Combining Eqs.~(\ref{eq:KSF-analytic}) and (\ref{eq:ZSF-analytic}), we get to the formula Eq.~(\ref{eq:ZtoNK-ratio}).

\nocite{*}

\bibliography{apssamp}% Produces the bibliography via BibTeX.

\end{document}